\documentclass[10pt]{article} 

\usepackage{arxiv}

\usepackage[utf8]{inputenc} 
\usepackage[T1]{fontenc}    
\usepackage{cite}
\usepackage{url}            
\usepackage{booktabs}       
\usepackage{amsfonts}       
\usepackage{nicefrac}       
\usepackage{microtype}      
\usepackage{lipsum}
\usepackage{graphicx}
\usepackage{multicol}
\graphicspath{ {./images/} }

\title{Large Language Models as Agents in the Clinic}

\author{
 Nikita Mehandru$^*$ \\
  UC Berkeley\\
  \texttt{nmehandru@berkeley.edu} \\
   \And
 Brenda Y. Miao$^*$ \\
  UCSF \\
  \texttt{brenda.miao@ucsf.edu} \\
\And
 Eduardo Rodriguez Almaraz$^*$ \\
  UCSF \\
  \texttt{eduardo.rodriguez@ucsf.edu} \\
\And
 Madhumita Sushil \\
  UCSF \\
  \texttt{madhumita.sushil@ucsf.edu} \\
\And
 Atul J. Butte \\
  UCSF \\
  \texttt{atul.butte@ucsf.edu} \\
\And
 Ahmed Alaa \\
  UC Berkeley \& UCSF \\
  \texttt{amalaa@berkeley.edu} \\
}

\begin{document}
\maketitle
\def\thefootnote{*}\footnotetext{equal contribution}\def\thefootnote{\arabic{footnote}}

\begin{abstract}
Recent developments in large language models (LLMs) have unlocked new opportunities for healthcare, from information synthesis to clinical decision support. These new LLMs are not just capable of modeling language, but can also act as intelligent “agents” that interact with stakeholders in open-ended conversations and even influence clinical decision-making. Rather than relying on benchmarks that measure a model’s ability to process clinical data or answer standardized test questions, LLM agents should be assessed for their performance on real-world clinical tasks. These new evaluation frameworks, which we call “Artificial-intelligence Structured Clinical Examinations” (“AI-SCI"), can draw from comparable technologies where machines operate with varying degrees of self-governance, such as self-driving cars. High-fidelity simulations may also be used to evaluate interactions between users and LLMs within a clinical workflow, or to model the dynamic interactions of multiple LLMs. Developing these robust, real-world clinical evaluations will be crucial towards deploying LLM agents into healthcare.
\end{abstract}

\vspace{12pt}
\begin{multicols}{2}
The release of ChatGPT, a chatbot powered by a large language model (LLM), has brought LLMs into the spotlight and unlocked new opportunities for their utilization in healthcare systems. Even though many of these large LLMs are trained on a variety of openly available information from the Internet rather than just biomedical information, these models have immense progress in clinical natural language processing (NLP) \cite{agrawal2022large,brown2020language,bubeck2023sparks} and have the potential to improve and augment clinical workflows.  For instance, the Generative Pretrained Transformer 4 (GPT-4) model can generate summaries of physician–patient encounters from transcripts of the conversation \cite{lee2023benefits}, achieve a score of 86\% on the United States Medical Licensing Examination (USMLE) \cite{fleming2023assessing}, and even create clinical question-answer pairs that are largely indistinguishable from human-generated USMLE questions \cite{nori2023capabilities}. These early demonstrations of GPT4 and other LLMs on clinical tasks and benchmarks suggest that these models have the potential to improve and automate clinical tasks.  

However, the emergent capabilities of LLMs have significantly expanded their potential applications beyond conventional, standardized clinical NLP tasks that primarily revolve around text processing and question answering. Instead, there is a growing emphasis on utilizing LLMs as Chatbots in both physician-facing and patient-facing tasks, where they are able to synthesize information, make medical inferences, or generate suggestions in a manner similar to a human expert  \cite{lee2023benefits,dash2023evaluation,wornow2023shaky}. In these scenarios, LLMs should not be viewed as models of language, but rather as intelligent “agents” that can interact with stakeholders in open-ended conversations and even influence clinical decision-making. eHealthcare systems are already adopting LLM-based chatbots; for instance, UC San Diego Health is already working to integrate GPT-4 into MyChart, Epic’s online health portal to streamline patient messaging \cite{DrChatbot}. Patients are also already leveraging publicly available chatbots (such as ChatGPT) to better understand medical vocabulary from clinical notes, and some medical centers are even considering exploring a “virtual-first” approach where LLMs assist in patient triaging \cite{levine2023diagnostic,korngiebel2021considering}. In all of these use cases, these LLMs “agents” go beyond traditional NLP tasks, and are instead used as active participants that contribute to clinical decision-making processes or engage in healthcare workflows in autonomous or semi-autonomous manners, both within and outside hospital settings. 

\section{Agent-based modeling of LLM chatbots}
\label{sec:headings}
To evaluate the utility and safety of LLM-based chatbots as agents in these novel and forthcoming applications, we suggest the use of novel benchmarks that are not confined to traditional, narrowly-scoped assessments based on NLP benchmarks consisting of predetermined inputs and ground-truths. Given that agency entails autonomy, it becomes imperative to evaluate LLMs in a manner similar to other comparable technologies where machines operate with varying degrees of self-governance.  

To evaluate LLMs effectively, one promising approach is to integrate concepts and tools from the domain of agent-based modeling (ABM) \cite{bankes2002agent}, a commonly used tool in health policy, complex systems, biology, ecology, economics, and the social sciences. ABM is a computational framework that enables simulation of the actions and interactions of autonomous agents, providing insights into system behavior and the factors influencing its outcomes. With an ABM approach, we can simulate multi-agent environments where LLMs interact with physicians, patients, and caretakers. Through these simulations, we can demonstrate and quantify emergent behavior, identify failure scenarios, and assess the impact of LLMs as an intervention in a healthcare system as well as other safety considerations. Guardrails, which have been developed for general-purpose models to constrain their behavior \cite{nvidia}, can also be developed for clinical LLMs based on insights derived from such simulations.  

An illustrative instance of applying ABM to evaluate a technology that demonstrates some level of agency can be found in the domain of self-driving cars \cite{fagnant2014travel}.  In this field, simulation environments that emulate the behavior of autonomous vehicles, drivers, and pedestrians are used to identify critical scenarios. A prominent example is Waymo, an autonomous driving technology company that utilizes agent-based simulation environments like “CarCraft” \cite{waymo1} and “Simulation City” \cite{waymo2} to thoroughly evaluate and refine their algorithms. Similar to standards and regulations for the autonomous driving industry, identifying robust clinical guidelines and what constitutes a successful interaction for healthcare LLMs will be crucial towards fulfilling the long-term goals of patients, providers, and other clinical stakeholders.

\section{Utility of agent-based modeling}
ABM approaches are already used in health research to conduct simulation studies of health behaviors, social epidemiology and the spread of infectious diseases \cite{tracy2018agent}. In all of these settings, ABM-based Monte Carlo simulations are used to thoroughly examine the effects of interactions among agents on system-level outcomes \cite{bonabeau2002agent}. Similarly, ABM-based approaches can be used to study the outcomes of LLM-Human interactions at scale, taking into consideration the probabilistic elements of both human and LLM behaviors.  

One aspect of LLMs that can be studied through ABM simulations is the impact of their sensitivity to user prompts on system outcomes. It is well-known that LLMs are highly influenced by the specific prompts given by users, such as physicians, healthcare workers, or patients \cite{lu2022fantastically}. In fact, understanding how to craft queries has become a significant subfield in computer science known as “prompt engineering”. An ABM can simulate the variability in language usage among different people, and a Monte Carlo approach can reveal the range of probable outcomes of an LLM-augmented system of care. Emerging phenomena that are caused by an LLM’s sensitivity to prompts can then be studied. For instance, disparities in outcomes arising from the varied quality of an LLM’s responses to patients of different backgrounds and native languages can be simulated to assess model fairness prior to its deployment.  

Agent-based simulations can also offer valuable insights into modeling interactions between users and LLMs, as well as the dynamic roles of multiple LLMs within the clinical workflow. Concerns surrounding the generation of inaccurate or biased information from LLMs have prompted researchers to explore approaches where models can cross-validate each other’s outputs to enhance consistency and accuracy \cite{dash2023evaluation,wang2022self}. Moreover, LLMs can be trained with specialized knowledge, such as ClinicalT5, which focuses exclusively on radiology reports and discharge summaries \cite{lehman2023clinical}. Complex clinical decision-making tasks, such as tumor boards, may involve the coordinated interaction of multiple LLMs with different areas of expertise. Another scenario could involve the collaboration between an LLM specializing in triage and another specializing in diagnosis. These examples highlight the interconnected nature of the interactions between humans and LLMs which contribute to shaping the trajectory of patient care. Utilizing an ABM approach allows for the simulation of numerous episodes involving these multi-agent interactions, enabling the evaluation of a distribution of possible outcomes under these interactions.

\section{Building an ABM simulation environment}
In order to evaluate LLMs using agent-based approaches, we need high-fidelity simulation environments of the healthcare systems in which these models will be deployed. Interestingly, LLMs themselves can be utilized in constructing these simulators. Previous research has demonstrated the feasibility of employing LLMs to create “interactive simulacra” that replicate human behavior \cite{park2023generative, yang2023auto}. By combining LLM-based models of patients and physicians with rule-based models of standard clinical workflows, it becomes possible to simulate the environments in which LLM models will be deployed. These simulation environments serve multiple purposes. Firstly, they can facilitate the evaluation of new LLM agents, allowing for rigorous assessments of their performance. Additionally, they can provide valuable insights into system-level aspects of the clinical workflow. Through these simulations, researchers can identify the specific types of information that are essential to extract and discuss in various complex clinical scenarios. Furthermore, systematic ablation experiments can be conducted to determine which information is not necessary for optimal outcomes. Moreover, the agents themselves can be treated as discrete components, systematically added or removed from interactions, to assess the contributions of specific roles to overall outcomes. This approach enables a comprehensive understanding of the impact of different agent configurations on the system.  

In human medical education, there has been a movement away from standardized testing that evaluate only shallow clinical reasoning and modern curricula increasingly use Objective Structured Clinical Examinations (OSCE) \cite{zayyan2011objective}. These exams assess a student’s practical skills in the clinic, including their ability to examine patients, take clinical histories, communicate effectively, and handle unexpected situations. Similarly, current benchmarks for clinical NLP, including MedQA (USMLE style questions) or MedNLI (identifies relationship between clinical sentences), are often also derived from standardized tests or curated clinical text and are not sufficient to capture the full range of capabilities demonstrated by clinical LLM agents \cite{lehman2023clinical,singhal2022large}. Instead, we call for the development of Artificial-intelligence Structured Clinical Examinations (“AI-SCI") that can be used to assess the ability for LLMs to aid in real-world clinical workflows. These AI-SCI benchmarks, which may be derived from difficult simulation scenarios or from real-world clinical tasks, should be created with input from interdisciplinary teams of clinicians, computer scientists, and the medical informatics community. As LLMs evolve and demonstrate increasingly advanced capabilities, their involvement in clinical practice will extend beyond limited text processing tasks. They will play a significant role in clinical decision-making and influence the cognitive load of healthcare professionals. In the near future, it may become necessary to shift our benchmarks from static datasets to dynamic simulation environments and transition from language modeling to agent modeling. Drawing inspiration from fields such as biology, the social sciences, and economics could be beneficial for future LLM research and development endeavors for clinical applications.
\end{multicols}

\bibliographystyle{unsrt}  
\bibliography{references}  


\end{document}